\documentclass[reprint,superscriptaddress,amsmath,amssymb,aps,longbibliography]{revtex4-2}
\usepackage{graphicx}
\usepackage{dcolumn}
\usepackage{bm}
\usepackage{hyperref}
\hypersetup{colorlinks=true,linkcolor=blue,urlcolor=blue,citecolor=blue}
\usepackage{float} 
\usepackage{subfigure} 
\usepackage{array}
\usepackage{booktabs} 
\usepackage{multirow}
\begin{document}

\title{Chebyshev pseudosite matrix product state approach for the spectral functions of electron-phonon coupling systems}
\author{Pei-Yuan Zhao}
\affiliation{State Key Laboratory of Low Dimensional Quantum Physics and Department of Physics, Tsinghua University, Beijing 100084, China}
\author{Ke Ding}
\affiliation{State Key Laboratory of Low Dimensional Quantum Physics and Department of Physics, Tsinghua University, Beijing 100084, China}
\author{Shuo Yang}
\email{shuoyang@tsinghua.edu.cn}
\affiliation{State Key Laboratory of Low Dimensional Quantum Physics and Department of Physics, Tsinghua University, Beijing 100084, China}
\affiliation{Frontier Science Center for Quantum Information, Beijing 100084, China}
\affiliation{Hefei National Laboratory, Hefei 230088, China}

\begin{abstract}
The electron-phonon ($e$-ph) coupling system often has a large number of phonon degrees of freedom, whose spectral functions are numerically difficult to compute using matrix product state (MPS) formalisms.
To solve this problem, we propose a new and practical method that combines the Chebyshev MPS and the pseudosite density matrix renormalization group (DMRG) algorithm.
The Chebyshev vector is represented by a pseudosite MPS with global $U(1)$ fermion symmetry, which maps $2^{N_p}$ bosonic degrees of freedom onto $N_p$ pseudosites, each with two states. 
This approach can handle arbitrary $e$-ph coupling Hamiltonians where pseudosite DMRG performs efficiently.
We use this method to study the spectral functions of the doped extended Hubbard-Holstein model in a regime of strong Coulomb repulsion, which has not been studied extensively before. 
Key features of the excitation spectra are captured at a modest computational cost.
Our results show that weak extended $e$-ph couplings can increase the spectral weight of the holon-folding branch at low phonon frequencies, in agreement with angle-resolved photoemission observations on one-dimensional (1D) cuprates.
\end{abstract}
\maketitle

\section{Introduction}

Electron-phonon couplings are of great interest due to the various phenomena they exhibit, such as Holstein polaron~\cite{PhysRevB.60.1633, PhysRevB.65.174306,PhysRevB.82.104304,PhysRevB.65.144301,PhysRevB.69.064302,PhysRevB.73.214304, Loos2006PhononSF, Loos2006SpectralFO,PhysRevB.55.14872,Paganelli_2006, PhysRevLett.114.146401,Chen2017,PhysRevB.68.184304,PhysRevB.71.115201,PhysRevB.100.094307,PhysRevB.102.165155}, pair density wave (PDW)~\cite{PhysRevLett.125.167001,huang2022pair}, and long-range attractive interactions in one-dimensional (1D) cuprate chains~\cite{chen2021anomalously,wang2021phonon}. 
A commonly studied model that incorporates local electron-phonon interactions is the 1D (extended) Hubbard-Holstein model (HHM)~\cite{fehske2004quantum, PhysRevLett.99.146404,Fehske_2008,PhysRevB.90.195134, Werner_2015,PhysRevB.64.094507, PhysRevB.63.035110,PhysRevB.89.014513,PhysRevB.75.245103,PhysRevB.76.155114,PhysRevB.99.075108,PhysRevLett.96.156402,hohenadler2013excitation,PhysRevLett.95.096401,PhysRevB.91.235150,PhysRevLett.109.116407}, which captures important features observed in numerous experiments.

Spectral functions such as local density of states (LDOS) and photoemission spectrum bridge the gap between theory and experiment. 
It can be challenging to calculate spectral functions for models with both strong electron-electron and electron-phonon interactions~\cite{PAECKEL2019167998,STOLPP2021108106,Ren2022}.
For example, several methods have been used to study the spectral functions of HHM, but each has its own limitations.
Exact diagonalization (ED)~\cite{fehske2004quantum} is limited to small system sizes due to the exponential wall problem and large phonon degrees of freedom. 
Cluster perturbation theory (CPT)~\cite{senechal2000spectral,senechal2002cluster,hohenadler2003spectral} with local basis optimization (LBO)~\cite{zhang1998density, zhang1999dynamical,weisse2000optimized} struggles to solve models with non-local interactions~\cite{zhao2005spectral,PhysRevLett.96.156402}. 
Additionally, it is difficult for Quantum Monte Carlo (QMC)~\cite{hohenadler2013excitation} to determine zero-temperature spectral functions.

Nowadays, density matrix renormalization group (DMRG)~\cite{PhysRevLett.69.2863, PhysRevB.48.10345,PhysRevB.72.180403} is the state-of-the-art numerical method for 1D strongly correlated systems. 
The matrix product state (MPS) representation of DMRG~\cite{PhysRevLett.75.3537,PhysRevLett.93.227205,PhysRevB.73.094423} makes this algorithm more flexible and easier to develop. 
Several variations of DMRG can be used to calculate the spectral functions of the (Hubbard-)Holstein model~\cite{PhysRevB.90.195134,Jiang2020,Jiang2021}, including dynamical DMRG (DDMRG)~\cite{PhysRevB.52.R9827, PhysRevB.60.335, PhysRevB.66.045114}, time-dependent DMRG (tDMRG) ~\cite{Daley_2004,PhysRevLett.93.076401,PhysRevLett.93.040502,PhysRevLett.107.070601,PhysRevB.94.165116, PAECKEL2019167998}, and Chebyshev MPS (CheMPS) method~\cite{PhysRevB.83.195115,PhysRevB.91.115144,PhysRevB.92.115130,PhysRevB.97.075111}. 
However, these methods are numerically expensive when dealing with systems with large local degrees of freedom. 
To reduce computational costs, Ref.~\cite{PhysRevB.74.241103} has combined DDMRG and pseudosite DMRG~\cite{PhysRevB.57.6376} to calculate the spectral functions of the half-filled HHM, which is accurate but time-consuming.
The tDMRG methods with optimized boson basis (OBB)~\cite{zhang1998density, zhang1999dynamical, weisse2000optimized, PhysRevLett.108.160401,PhysRevB.93.075105,PhysRevB.92.241106} and the projected purification method~\cite{Kohler:2020hkc} have also been developed for spectral functions~\cite{PhysRevB.102.165155,Mardazad2021}. 
These approaches can truncate the phonon basis with a small loss of accuracy but are difficult to employ. 
In addition, the entanglement entropy of MPS increases quickly with time, making it costly to calculate states over a long time interval.

In this article, we provide a simple and computationally efficient method for calculating the spectral functions of strongly correlated systems with extended $e$-ph couplings. 
We combine the $U(1)$-symmetric pseudosite MPS (SPS-MPS) formalism with the Chebyshev MPS approach for the first time.
This method is used to investigate the (anti-)photoemission spectral functions of the doped extended HHM with large Hubbard $U$, which has not been thoroughly explored previously.

For SPS-MPS, the fermionic and bosonic degrees of freedom of a lattice site are replaced by multiple pseudosites, where a real boson with $2^{N_p}$ truncated states is mapped to $N_p$ hard-core bosons, similar to the pseudosite DMRG method~\cite{PhysRevB.57.6376}. 
The transformation from the original Hamiltonian to the pseudosite Hamiltonian involves reshaping and decomposing the matrix product operator (MPO), which can be easily done in the MPO-MPS representation. 
The global $U(1)$ symmetry of fermions is encoded in the pseudosite tensor networks, where all MPOs and MPSs are in block-sparse forms~\cite{McCulloch_2007,PhysRevA.82.050301,PhysRevB.83.115125}. 

The Chebyshev pseudosite MPS approach is used to calculate the zero-temperature spectral functions of an electron-phonon coupling system
\begin{align}
G(\omega)=\langle \psi_0|\hat{O}^{\dagger}\delta(\omega-\hat{H}+E_0)\hat{O}|\psi_0\rangle,
\label{Eq1}
\end{align}
where $\hat{O}$ is a certain operator, $\hat{H}$ is the Hamiltonian, and $|\psi_0\rangle$ is the ground state with eigenenergy $E_0$. 
The spectral function Eq.~(\ref{Eq1}) is expanded by Chebyshev polynomials. 
In Chebyshev iterations, all states and operators are represented as $U(1)$-symmetric pseudosite MPSs and MPOs. 
Our method can calculate the spectral functions of an arbitrary $e$-ph coupling Hamiltonian as long as pseudosite DMRG can efficiently compute its ground state.

The extended HHM (eHHM)~\cite{wang2021phonon} is defined as
\begin{align}
\hat{H}=&-t\sum_{i,\sigma}(\hat{c}_{i,\sigma}^{\dagger}\hat{c}_{i+1,\sigma}+h.c.)+U\sum_i\hat{n}_{i\uparrow}\hat{n}_{i\downarrow}-\mu\sum_i\hat{n}_i\nonumber \\
&+\omega_0\sum_{j}\hat{a}_j^{\dagger}\hat{a}_j+\sum_{\{ij\},\sigma}\gamma_{ij}(\hat{a}_{i}^{\dagger}+\hat{a}_i)\hat{n}_{j,\sigma},\label{Eq2}
\end{align}
where $\hat{c}_{i,\sigma}^{\dagger}$ and $\hat{c}_{i,\sigma}$ ($\hat{a}_i^{\dagger}$ and $\hat{a}_i$) are electron (phonon) creation and annihilation operators, and $\hat{n}_{i}=\sum_{\sigma}\hat{c}_{i,\sigma}^{\dagger}\hat{c}_{i,\sigma}$ is the electron number operator. 
We consider on-site, nearest-neighbor (NN), and next-nearest-neighbor (NNN) electron-phonon couplings ($\gamma$, $\gamma'$, and $\gamma''$). 
When $\gamma'=\gamma''=0$, eHHM reduces to HHM.

We first calculate the Green's function of the Holstein polaron (HP) and the (anti-)photoemission spectral functions of the half-filled HHM in the Mott insulator phase to serve as a benchmark. 
Our method can reproduce key features in the excitation spectra of the (Hubbard-)Holstein model. 
Then, inspired by recent experiments with 1D cuprates~\cite{chen2021anomalously,wang2021phonon}, we investigate the effects of extended $e$-ph couplings by computing spectral functions of doped eHHM with strong Coulomb repulsion ($U/t=8$). 
The Hubbard $U$ and the ratios between $e$-ph coupling constants ($\gamma'=\gamma/\sqrt{7}, \gamma''=\gamma/\sqrt{15}$) are chosen according to 1D cuprates~\cite{chen2021anomalously,wang2021phonon}. 
When the phonon frequency is comparable to electron hopping ($\omega_0/t=1$), we find that the charge band is no longer connected to the spinon band and its spectral weight is reduced by phonon effects.
Furthermore, in the adiabatic limit ($\omega_0/t=0.2$) and for lightly doped cases, we observe that the spectral weight of the holon-folding (hf) branch is enhanced while the $3k_F$ branch is suppressed, which agrees with ARPES experiments~\cite{chen2021anomalously}.
Our findings demonstrate that even when the on-site $e$-ph coupling is weak, extended $e$-ph couplings ($\gamma', \gamma''$) can have a significant influence on spectral functions, which may be important for understanding the nature of high-$T_c$ superconductors.

This article is organized as follows. 
In Sec.~\ref{section2A}, we introduce the $U(1)$-symmetric pseudosite MPO-MPS formalism. 
In Sec.~\ref{section2B}, we present the Chebyshev iteration method using SPS-MPS and SPS-MPO.
In Sec.~\ref{section3}, we show numerical results for the spectral functions of the (Hubbard-)Holstein model and the doped eHHM. 
Finally, in Sec.~\ref{section4}, we summarize our findings and provide an outlook for future research.

\section{Methods}

\subsection{$U(1)$-symmetric pseudosite MPO-MPS formalism}
\label{section2A}

\begin{figure}
\centering 
\includegraphics[width=1.0\columnwidth]{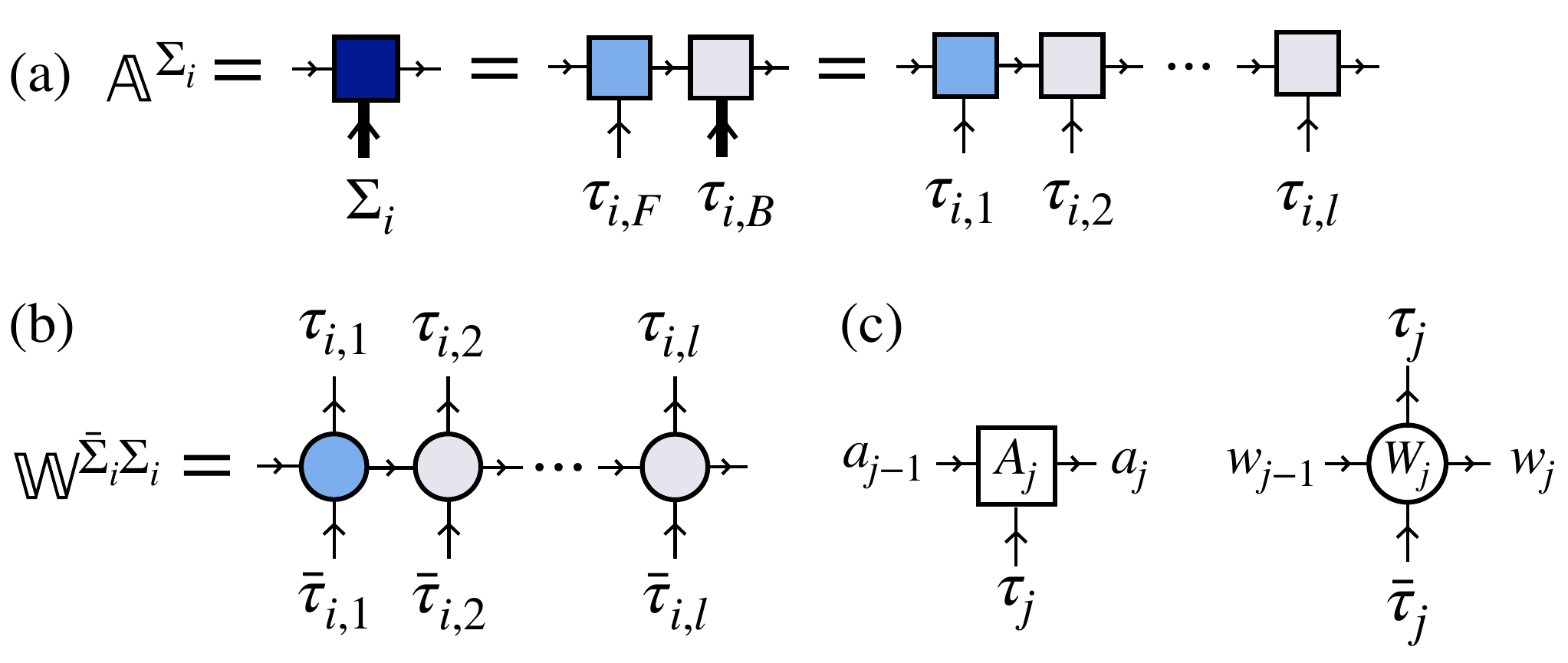} 
\caption{
(a) The MPS on the $i$-th lattice site is decomposed into fermionic and bosonic components, which are further decomposed into $l$ pseudosite MPSs.
(b) The MPO on the $i$-th lattice site is decomposed into $l$ pseudosite MPOs.
(c) MPS and MPO on the $j$-th pseudosite. 
} 
\label{Fig.main1} 
\end{figure}

We start from the usual MPO-MPS formalism, where an $N$-site Hamiltonian and variational wave functions are represented by MPO and MPS, respectively.
\begin{align}
\hat{H}[\mathbb{W}]&=\sum_{\bar{\boldsymbol{\tau}},\boldsymbol{\tau}}(\mathbb{W}^{\bar{\Sigma}_0\Sigma_0}\mathbb{W}^{\bar{\Sigma}_1\Sigma_1}\cdots \mathbb{W}^{\bar{\Sigma}_{N-1}\Sigma_{N-1}})|\bar{\boldsymbol{\tau}}\rangle\langle\boldsymbol{\tau}|, \nonumber\\
|\Psi[\mathbb{A}]\rangle &=\sum_{\boldsymbol \tau}(\mathbb{A}^{\Sigma_0}\mathbb{A}^{\Sigma_1}\cdots\mathbb{A}^{\Sigma_{N-1}})|\boldsymbol{\tau}\rangle,
\end{align}
where $\mathbb{A}^{\Sigma_i}$ and $\mathbb{W}^{\bar{\Sigma}_{i},\Sigma_i}$ are MPS and MPO on the $i$-th lattice site, and $|\boldsymbol{\tau} \rangle$ is the basis of the Hilbert space.

As illustrated in Fig.~\ref{Fig.main1}(a), $\mathbb{A}^{\Sigma_i}$ is first divided into fermionic and bosonic components
\begin{align}
\mathbb{A}^{\Sigma_i}= A^{\tau_{i,F}}A^{\tau_{i,B}}.
\end{align}
Here $\tau_{i,F}$ and $\tau_{i,B}$ denote the fermionic index with $4$ states and the bosonic index with $2^{N_p}$ states. 
Next, we reshape $\tau_{i,B}$ to generate ${N_p}$ pseudosite indexes~\cite{PhysRevB.57.6376} 
\begin{align}
\tau_{i,B}=(\tau_{i,2},\tau_{i,3},\cdots, \tau_{i,l}),
\label{Eq5}
\end{align}
where each index $\tau_{i,m}$ only contains $2$ states. 
This corresponds to mapping one real boson to $N_p$ hard-core bosons.
As a result, we obtain an array of pseudosite MPS with length $l=N_p+1$ by denoting $\tau_{i,F}$ as $\tau_{i,1}$ [see Fig.~\ref{Fig.main1}(a)]
\begin{align}
\mathbb{A}^{\Sigma_i}=A^{\tau_{i,1}}\cdots A^{\tau_{i,l}}.
\label{Eq6}
\end{align}
Similarly, $\mathbb{W}^{\bar{\Sigma}_{i},\Sigma_i}$ is decomposed into an array of pseudosite MPO [see Fig.~\ref{Fig.main1}(b)]
\begin{align}
\mathbb{W}^{\bar{\Sigma}_{i},\Sigma_i}=W^{\bar{\tau}_{i,1}\tau_{i,1}}\cdots W^{\bar{\tau}_{i,l}\tau_{i,l}}.
\end{align}
To make the notation more convenient, we rename the $(i,m)$-th pseudosite MPS (MPO) by $j=li+m$, i.e., $A_j=A^{\tau_{i,m}}$ and $W_j=W^{\bar{\tau}_{i,m}\tau_{i,m}}$, where $1\leq m\leq l$. 
Fig.~\ref{Fig.main1}(c) depicts the $j$-th pseudosite MPS $A^{\tau_j}\in \mathbb{C}^{D_{j-1}\times d_j\times D_j}$ and MPO $W^{\bar{\tau}_j,\tau_j}\in \mathbb{C}^{\chi_{j-1}\times d_j\times \chi_j\times d_j}$, where $D_j$ and $\chi_j$ are the virtual bond dimensions and $d_j$ is the physical bond dimension.

In practice, we usually start with a random pseudosite MPS as the initial variational wave function, unless otherwise specified. 
The procedure depicted in Fig.~\ref{Fig.main1}(b) is carried out through successive singular value decompositions (SVD)~\cite{SCHOLLWOCK201196,PhysRevA.81.062337,PhysRevB.95.035129}, during which singular values less than $10^{-7}$ are ignored for simplicity.
These truncations may introduce errors, which are estimated by $\varepsilon_{\mathrm{local}}$ and  $\varepsilon_{\mathrm{global}}$,
\begin{align}
\varepsilon_{\mathrm{local}}&=\ \Vert W^{\bar{\tau}_{i,B},\tau_{i,B}}-W^{\bar{\tau}_{i,2},\tau_{i,2}}\cdots W^{\bar{\tau}_{i,l},\tau_{i,l}}\Vert,\nonumber\\
\varepsilon_{\mathrm{global}}&=\Vert \hat{H}-\hat{H}_{\mathrm{pseu}}\Vert/\sqrt{\Vert \hat{H}\Vert\Vert \hat{H}_{\mathrm{pseu}}\Vert}.
\end{align}
$\varepsilon_{\mathrm{local}}$ is the $2$-norm of the difference between the actual bosonic MPO and the product of corresponding pseudosite MPOs. 
$\varepsilon_{\mathrm{global}}$ is the relative $2$-norm distance between the exact Hamiltonian $\hat{H}$ and the Hamiltonian $\hat{H}_{\mathrm{pseu}}$ represented by pseudosite MPOs.
Table~\ref{Table1} shows the relation between $\{\chi_j\}$ and $N_p$ for Hamiltonian Eq.~(\ref{Eq2}), where $\chi_j$ only slowly increases with $N_p$.
\begin{table}
\centering
\begin{ruledtabular}
\begin{tabular}{ccc}
$N_p$ & Model & $\{\chi_{j}\}$  \\ 
\specialrule{0.05em}{3pt}{3pt}
3& HHM &\{7,10,8,7\} \\
4& HHM &\{7,10,11,8,7\} \\
5& HHM &\{7,10,12,11,8,7\} \\
6& HHM &\{7,10,12,14,11,8,7\} \\
7 & HHM &\{7,10,12,14,14,11,8,7\} \\
3 & HHM+$\gamma'$ &\{9,14,11,9\}\\
4 & HHM+$\gamma'$ &\{9,14,17,11,9\}\\
5 & HHM+$\gamma'$ &\{9,14,18,17,11,9\}\\
6 & HHM+$\gamma'$ &\{9,14,18,22,17,11,9\}\\
3 & HHM+$\gamma'$+$\gamma''$&\{12,16,13,12\}\\
4 & HHM+$\gamma'$+$\gamma''$&\{12,16,19,13,12\}\\
5 & HHM+$\gamma'$+$\gamma''$&\{12,16,20,19,13,12\}\\
6 & HHM+$\gamma'$+$\gamma''$&\{12,16,20,23,19,13,12\}\\
\end{tabular}
\end{ruledtabular}
\caption{
The virtual bond dimensions of the pseudosite MPOs generated by SVD.
$\gamma'$ and $\gamma''$ are the nearest-neighbor and next-nearest-neighbor $e$-ph coupling constants.
}
\label{Table1}
\end{table}
Table~\ref{Table2} shows that both $\varepsilon_{\mathrm{local}}$ and  $\varepsilon_{\mathrm{global}}$ are less than $10^{-7}$, indicating SVD truncation errors are well-controlled. 

By taking advantage of the $U(1)$ fermionic symmetry of the Hamiltonian Eq.~(\ref{Eq2}), block-sparse MPO-MPS formalisms~\cite{McCulloch_2007,PhysRevA.82.050301,PhysRevB.83.115125} are used to reduce the computational cost.
Each index of MPS/MPO carries a $U(1)$ charge of fermions and flows from site to site as indicated by arrows in Fig.~\ref{Fig.main1}(c). 
The MPS/MPO blocks have non-zero elements only when the inflowing and outflowing quantum numbers are conserved.
We note that only the fermionic pseudosite MPS $A_{j=li+1}$ contributes to the $U(1)$ charge for the $i$-th lattice site. 
Accordingly, the allowed quantum numbers for the physical index of $A_{j=li+1}$ are different from the ones of other local MPS $A_{j=\mathrm{others}}$,
\begin{align}
Q(\tau_{j=li+1})&=\left\{
\begin{aligned}
&0,\ \ \mathrm{if}\ |\tau_j\rangle=|\mathrm{vac}\rangle,  \\
&1,\ \ \mathrm{if}\ |\tau_j\rangle =c_{i,\uparrow}^{\dagger}|\mathrm{vac}\rangle \ \mathrm{or}\  c_{i,\downarrow}^{\dagger}|\mathrm{vac}\rangle,\\
&2,\ \ \mathrm{if}\ |\tau_j\rangle=c_{i,\uparrow}^{\dagger}c_{i,\downarrow}^{\dagger}|\mathrm{vac}\rangle,\\
\end{aligned}
\right.\nonumber \\
Q(\tau_{j=\mathrm{others}})&=0,
\label{Eq11}
\end{align}
where $Q(x_j)$ is the quantum number of a given index $x_j$ and $|\mathrm{vac}\rangle$ is the vacuum state. 
Because of the $U(1)$-invariance, the quantum numbers of the non-zero MPS/MPO blocks satisfy
\begin{align}
Q(a_{j-1})+Q(\tau_j)&=Q(a_j),\nonumber\\
Q(w_{j-1})+Q(\bar{\tau}_j)&=Q(\tau_j)+Q(w_j).
\end{align}
The quantum numbers of the leftmost and rightmost virtual indexes in an open chain with $N_e$ electrons are set to be zero and $N_e$, respectively, so that only the bases $\{|\boldsymbol{\tau}\rangle\}$ satisfying $\langle \boldsymbol{\tau}|\sum_i\hat{n}_i| \boldsymbol{\tau}\rangle=N_e$ contribute to the overall state.

\begin{table}
\centering
\begin{ruledtabular}
\begin{tabular}{ccc}
($N$,$U$,$N_p$,$\omega_0$,$\gamma$) &$\varepsilon_{\mathrm{global}}$ & $\varepsilon_{\mathrm{local}}$  \\ 
\specialrule{0.05em}{3pt}{3pt}
(16,8,3,1,1)& $4.71\times 10^{-8}$ & $6.82\times 10^{-15}$\\
(32,8,3,1,1)& $6.04\times 10^{-8}$& $6.82\times 10^{-15}$\\
(16,8,3,1,0.5)&$3.91\times 10^{-8}$ & $1.87\times 10^{-13}$ \\
(32,8,3,1,0.5)&$3.94\times 10^{-8}$ & $1.87\times 10^{-13}$ \\
(16,8,4,1,0.5)+$\gamma'$&$1.50\times 10^{-8}$ & $1.74\times 10^{-13}$ \\
(32,8,4,1,0.5)+$\gamma'$&$4.29\times 10^{-8}$ & $1.74\times 10^{-13}$ \\
(16,8,4,1,0.5)+$\gamma'$+$\gamma''$&$2.46\times 10^{-8}$ & $1.79\times 10^{-13}$ \\
(32,8,4,1,0.5)+$\gamma'$+$\gamma''$&$3.92\times 10^{-8}$ & $1.79\times 10^{-13}$ \\
\end{tabular}
\end{ruledtabular}
\caption{
The local and global truncation errors of pseudosite MPOs for various parameters ($N$, $U$, $N_p$, $\omega_0$, and $\gamma$).
}
\label{Table2}
\end{table}

\subsection{Chebyshev pseudosite MPS method}
\label{section2B}
We now introduce the Chebyshev pseudosite MPS method step by step. 
First, the standard DMRG~\cite{PhysRevLett.69.2863, PhysRevB.48.10345,PhysRevB.72.180403,PhysRevB.57.6376} is used to compute the ground state $|\psi_0\rangle$ with an eigenenergy $E_0$.
After that, the spectral function Eq.~(\ref{Eq1}) is expanded by the Chebyshev polynomials~\cite{RevModPhys.78.275}
\begin{align}
T_n(\omega'), \ -1\leq\omega'\leq 1.
\label{Eq14}
\end{align}
The domain of the spectral function Eq.~(\ref{Eq1}) is $\omega\in[0, E_{\mathrm{max}}-E_0]$, where $E_{\mathrm{max}}$ is the maximal eigenvalue of the Hamiltonian. 
While $E_{\mathrm{max}}$ scales with system size, the non-zero spectral weight can only be found in $\omega\in[0,\mathcal{W}_A]$, where $\mathcal{W}_A$ is the spectral width. 
In practice, we use the energy window $\omega\in[-\omega_{2\mathrm{max}},\omega_{1\mathrm{max}}]$, where $\omega_{1\mathrm{max}}$ is set to $2\mathcal{W}_A\sim3\mathcal{W}_A$, and $\omega_{2\mathrm{max}}$ is chosen depending on the model.
 
Since the proper domain of $T_n(\omega')$ is $[-1,1]$, we map the energy window to this range.
\begin{align}
\omega\mapsto\omega',\ \omega'=\frac{\omega}{a}+b,\ \omega'\in[-W',W'],
\end{align}
where
\begin{align}
a=\frac{\omega_{1\mathrm{max}}+\omega_{2\mathrm{max}}}{2W'},\ b=\frac{\omega_{2\mathrm{max}}-\omega_{1\mathrm{max}}}{\omega_{2\mathrm{max}}+\omega_{1\mathrm{max}}}W',
\end{align}
and $W'$ is a number slightly smaller than $1$. 
Correspondingly, the Hamiltonian is rescaled and shifted.
\begin{align}
\hat{H}\mapsto\hat{H}',\ \hat{H}'=\frac{\hat{H}-E_0}{a}+b,\ E'\in[b,\frac{E_{\mathrm{max}}-E_0}{a}+b].
\label{Eq16}
\end{align}

The spectral function is further represented as
\begin{align}
G(\omega)&=\langle \psi_0|\hat{O}^{\dagger}\delta(\omega+E_0-\hat{H})\hat{O}|\psi_0\rangle\nonumber\\
&=\frac{1}{a}\langle \psi_0|\hat{O}^{\dagger}\delta(\omega'-\hat{H}')\hat{O}|\psi_0\rangle\nonumber\\
&=G(\omega').
\end{align}
The first $N_{\mathrm{C}}$ terms of $T_n(\omega')$ are employed to expand $G(\omega')$ approximately,
\begin{align}
\delta(\omega'-\hat{H}')=\frac{1}{\pi\sqrt{1-\omega'^2}}[g_0+2\sum_{n=1}^{N_{\mathrm{C}}-1}g_nT_n(\hat{H}')T_n(\omega')],
\end{align}
\begin{align}
G(\omega')=\frac{1}{a\pi\sqrt{1-\omega'^2}}[g_0\mu_0+2\sum_{n=1}^{N_{\mathrm{C}}-1}g_n\mu_nT_n(\omega')],
\label{Eq19}
\end{align}
where
\begin{align}
\mu_n=\langle \psi_0|\hat{O}^{\dagger}T_n(\hat{H}')\hat{O}|\psi_0\rangle
\label{Eq20}
\end{align}
are the Chebyshev moments.
The damping factors $g_n$ are used to suppress the Gibbs oscillations~\cite{RevModPhys.78.275} caused by the truncation of the Chebyshev series. 
In this study, we use the Jackson damping
\begin{align}
g_n=\frac{(N_{\mathrm{C}}-n+1)\mathrm{cos}\frac{\pi n}{N_{\mathrm{C}}+1}+\mathrm{sin}\frac{\pi n}{N_{\mathrm{C}}+1}\mathrm{cot}\frac{\pi}{N_{\mathrm{C}}+1}}{N_{\mathrm{C}}+1}.
\end{align}
Since the Chebyshev polynomials $T_n(\omega')$ are known, once we have calculated the expansion coefficients $\mu_n$ in Eq.~(\ref{Eq20}), the spectral function Eq.~(\ref{Eq19}) will be obtained. 
$\mu_n$ can be computed by performing Chebyshev iterations
\begin{align}
|t_0\rangle&=\hat{O}|\psi_0\rangle,\ \ \ \ |t_1\rangle=\hat{H}'|t_0\rangle,\nonumber\\
|t_{n+1}\rangle&=2\hat{H}'|t_n\rangle-|t_{n-1}\rangle,
\label{Eq22}
\end{align}
where $|t_n\rangle$ are Chebyshev vectors and
\begin{align}
\mu_n=\langle t_0|t_n\rangle.
\end{align}

The focus of this article is on the photoemission and antiphotoemission spectral functions of the eHHM with a fixed number of fermions,
\begin{align}
A^{(-)}(k,\omega)=\langle \psi_0|\hat{c}_{k,\sigma}^{\dagger}\delta(-\omega+E_0-\hat{H})\hat{c}_{k,\sigma}|\psi_0\rangle,
\label{Eq24}
\end{align}
\begin{align}
A^{(+)}(k,\omega)=\langle \psi_0|\hat{c}_{k,\sigma}\delta(\omega+E_0-\hat{H})\hat{c}_{k,\sigma}^{\dagger}|\psi_0\rangle.
\label{Eq25}
\end{align}
We refer to $A(k,\omega)$ as the sum of photoemission (\ref{Eq24}) and antiphotoemission (\ref{Eq25}) spectral functions, which satisfies the normalization condition
\begin{align}
A(k,\omega)=A^{(-)}(k,\omega)+A^{(+)}(k,\omega), \int A(k,\omega)d\omega=1.
\end{align}
The Fourier transformations of fermionic creation and annihilation operators are introduced for open boundary conditions~\cite{PhysRevLett.92.256401}
\begin{align}
\hat{c}_{k,\sigma}^{\dagger}&=\sqrt{\frac{2}{L+1}}\sum_{i=0}^{N-1}\sin [k(i+1)] \hat{c}_{i,\sigma}^{\dagger},\nonumber\\
\hat{c}_{k,\sigma}&=\sqrt{\frac{2}{L+1}}\sum_{i=0}^{N-1}\sin [k(i+1)] \hat{c}_{i,\sigma}.
\label{Eq27}
\end{align}
and the (quasi-)momentum is
\begin{align}
k=\frac{\pi z}{N+1},\ 1\leq z\leq N.
\end{align}
The initial Chebyshev iteration $|t_0\rangle = \hat{c}^{\dagger}_{k,\sigma} | \psi_0 \rangle$ (or $|t_0\rangle = \hat{c}_{k,\sigma} | \psi_0 \rangle$) shifts the total number of fermions from $N_e$ to $N_e+1$ (or $N_e-1$), whereas all subsequent iterations do not.    
By variationally minimizing the fitting errors
\begin{align}
\Delta_{\mathrm{fit}}=\left\{
\begin{aligned}
&\Vert|t_{n}\rangle-\hat{H}'|t_{n-1}\rangle\Vert^2,\ \ \mathrm{if} \ n =1  \\
&\Vert|t_{n}\rangle-(2\hat{H}'|t_{n-1}\rangle-|t_{n-2}\rangle)\Vert^2,\ \ \mathrm{if} \ n \geq 2\\
\end{aligned}
\right.
\end{align}
using the two-site update method, we obtain $|t_n\rangle\ ( n\geq 1)$ with a fixed particle number $N_e\pm1$.
The computational cost for each step is $\mathcal{O}(LD^3d^2\chi)$, where $L$ is the length of the 1D tensor network, $D$ ($\chi$) is the maximal bond dimension of MPS (MPO), and $d$ is the physical degrees of freedom per pseudosite, i.e., $d=4$ for fermions and $d=2$ for hard-core bosons. 
Since the computational cost only scales linearly with $L$, decomposing a large physical degree of freedom into multiple pseudosites can save a lot of effort.

In Eq.~(\ref{Eq16}), the maximum eigenvalue of $H'$ is larger than $1$.
As a result, the high energy components of $|t_n\rangle$ must be projected out to avoid divergence, for which we use the energy truncation procedure described in Ref.~\cite{PhysRevB.83.195115}.
With the single-site update method, the computational cost per site is $\mathcal{O}(d_K^2D^3d\chi)$, where $d_K$ is the dimension of the local Krylov subspace. 
The truncation-induced state change is given by~\cite{PhysRevB.83.195115}
\begin{align}
\Delta_{\mathrm{tr}}=\Vert |t_n\rangle_{\mathrm{tr}}-|t_n\rangle\Vert^2,
\end{align}
where $|t_n\rangle_{\mathrm{tr}}$ is the truncated Chebyshev vector.

\section{Results}
\label{section3}

\subsection{Benchmarks}
\begin{figure}[tbp]
\centering
\includegraphics[width=1.0\columnwidth]{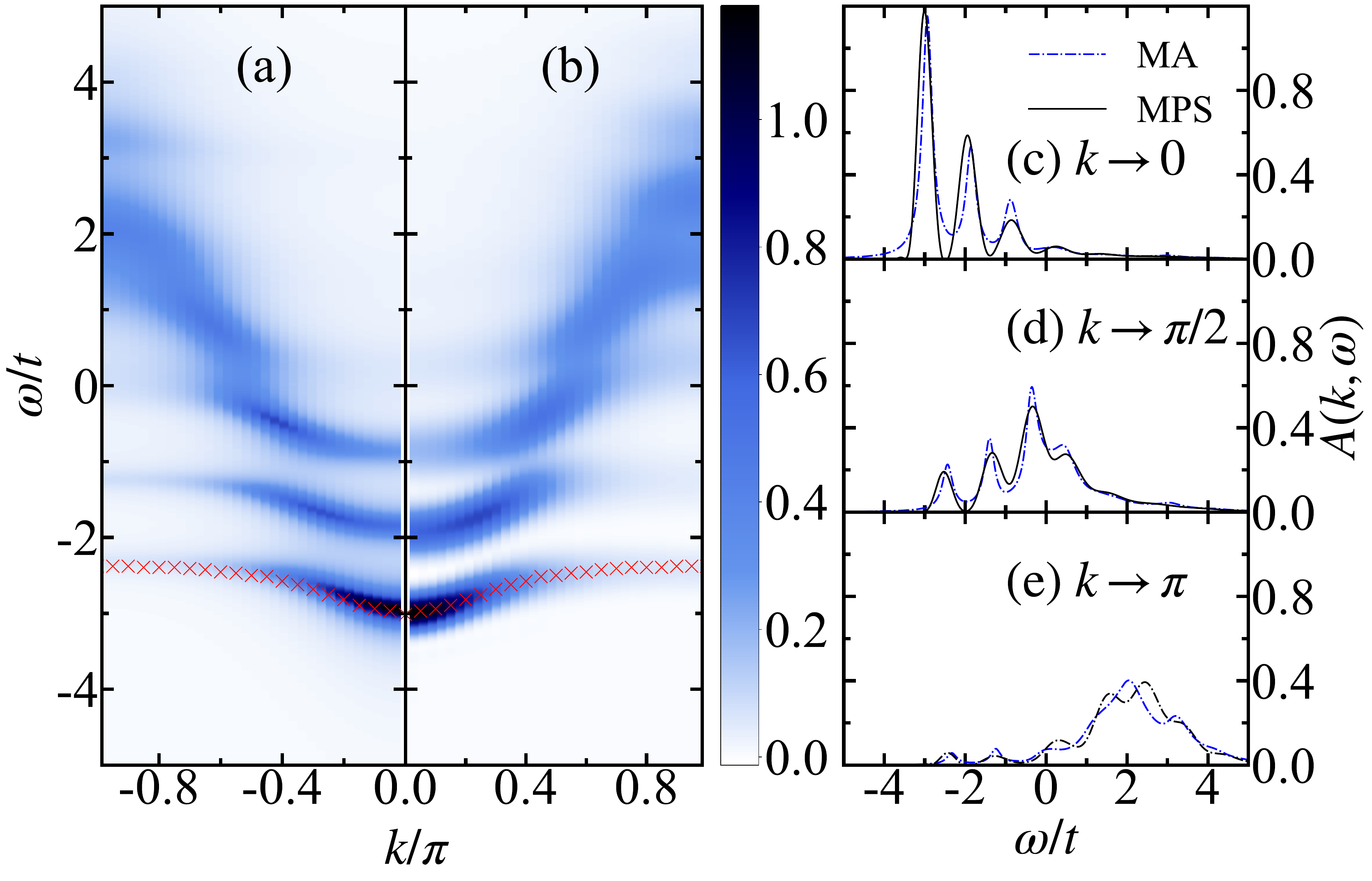}
\caption{
The spectral function of the Holstein polaron, where $\omega_0/t=1$, $\gamma/t=\sqrt{2}$, and $U/t=0$. 
(a) The spectrum given by momentum average approximation (MA) with $\eta=0.15$. 
(b) The spectral function calculated by SPS-MPS, where $N=32$, $N_p=4$, $N_{\mathrm{C}}=100$, $D=50$, $d_K=30$, $\omega_{1\mathrm{max}}=16$, and $\omega_{2\mathrm{max}}=0$.
Markers label the results from the variational method in Ref.~\cite{PhysRevB.60.1633}. 
(c-e) Comparisons of the spectral functions calculated by MA and SPS-MPS at different momenta $k$.
} 
\label{Fig.main2} 
\end{figure}
\begin{figure}[tbp]
\centering 
\includegraphics[width=1.0\columnwidth]{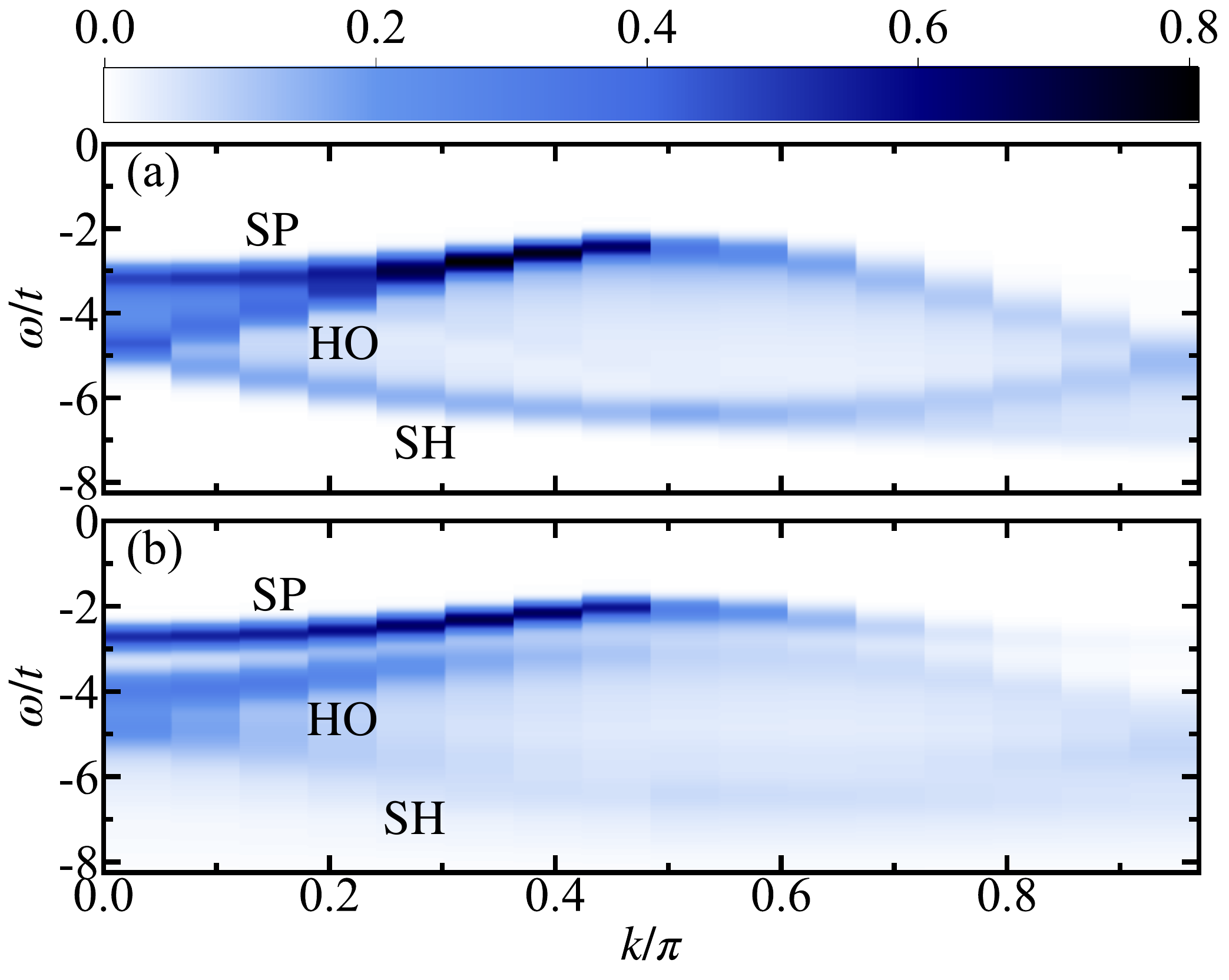} 
\caption{The spectral functions of the Mott insulators with $N=32$, $N_e=32$, $U/t=8$, $N_{\mathrm{C}}=100$, $d_K=30$, $\omega_{1\mathrm{max}}=15$, and $\omega_{2\mathrm{max}}=0$.
SP, HO, and SH represent the spinon band, the holon band, and the shadow band, respectively.
(a) The spectral functions of HM with $D=100$. (b) The spectral functions of HHM with $N_p=3$, $\omega_0/t=1$, $\gamma/t=1$, and $D=200$.}
\label{Fig.main3} 
\end{figure}

We conduct several tests to evaluate the performance of our method.
In particular, we calculate the Green's function of the Holstein polaron and the spectral functions of the half-filled HHM in the Mott insulator phase.

The Green's function of the Holstein polaron is
\begin{align}
A(k,\omega)=\langle \mathrm{vac}|\hat{c}_{k,\sigma}\delta(\omega-\hat{H})\hat{c}_{k,\sigma}^{\dagger}|\mathrm{vac}\rangle.
\label{Eq33}
\end{align}
Here, $\hat{H}$ is the Hamiltonian of the Holstein model with $U=0$ and only on-site $e$-ph couplings.
We choose a phonon frequency comparable to electron hopping ($\omega_0/t=1$) and use the $e$-ph coupling $\gamma/t=\sqrt{2}$, which lies in the crossover regime between the large ($\gamma/t<\sqrt{2}$) and small ($\gamma/t>\sqrt{2}$) polarons. 
As shown in Fig.~\ref{Fig.main2}(b), a couple of subbands correspond to the excitations of an electron with one or more phonons, the intervals between which are about $\omega_0$.
The polaron band agrees quantitatively with those from variational methods~\cite{PhysRevB.60.1633}, as indicated by the markers in Fig.~\ref{Fig.main2}(b). 
The bottom of the energy band is about $-3.01t$ when $k\rightarrow 0$, close to the variational result for an infinite lattice (see Table~\uppercase\expandafter{\romannumeral1} of Ref.~\cite{PhysRevB.60.1633}).
Moreover, the spectral functions calculated by SPS-MPS agree well with those from the momentum average approximation (MA)~\cite{PhysRevB.74.245104} with a finite broadening $\eta=0.15$, as demonstrated side by side in Fig.~\ref{Fig.main2}(a-b).

\begin{figure}[tbp]
\centering 
\includegraphics[width=1.0\columnwidth]{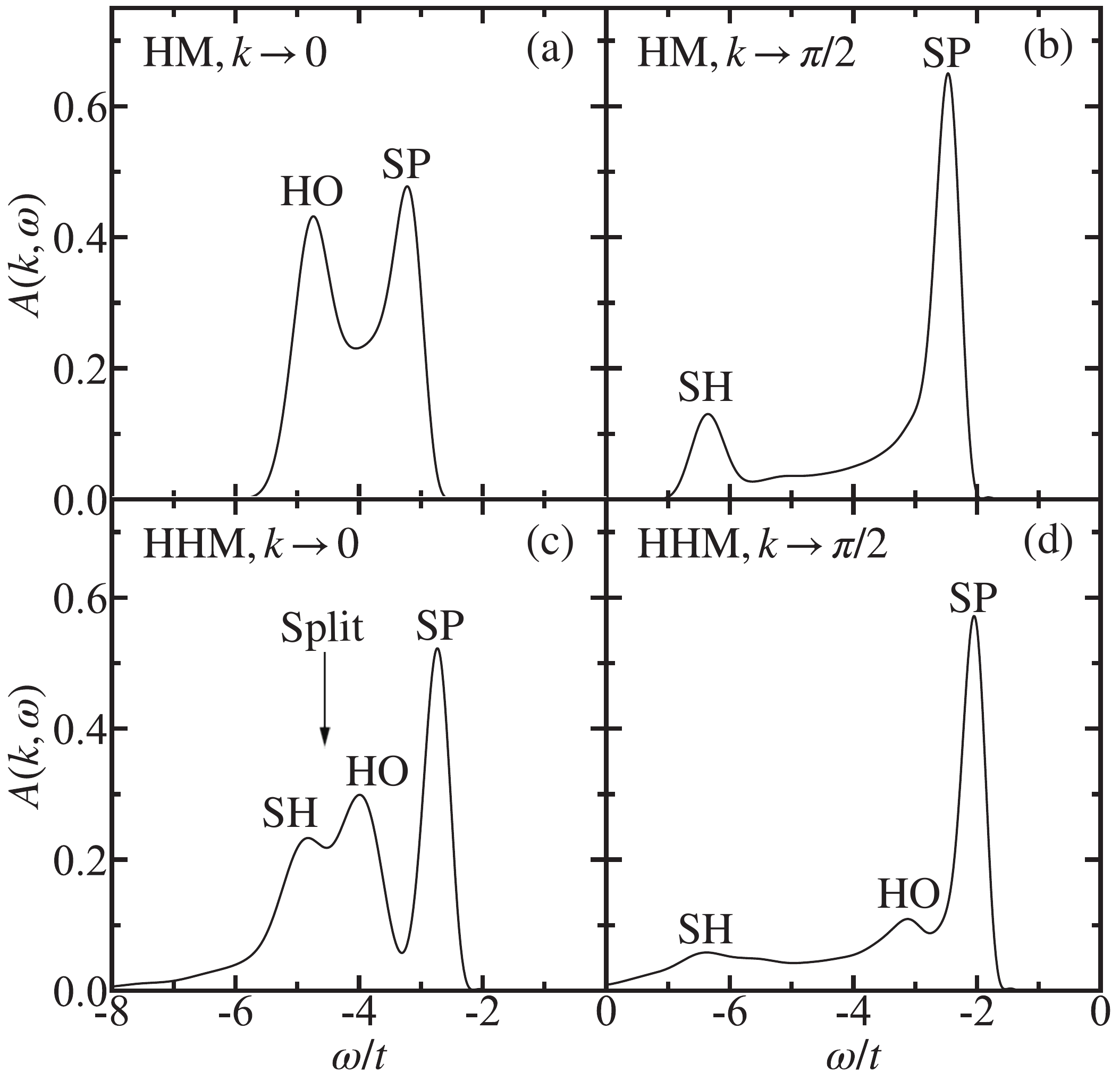} 
\caption{
The spectral functions of HM and HHM in the Mott insulator phase with fixed momentum $k$, where all parameters are the same as those in Fig.~\ref{Fig.main3}.
} 
\label{Fig.main4} 
\end{figure}

Next, we investigate the spectral functions of the Mott insulators with on-site $e$-ph couplings.
The spectral function of the half-filled Hubbard model (HM) is shown in Fig.~\ref{Fig.main3}(a), where the strong Coulomb repulsion causes a large Mott gap around $\omega=0$ and leads to spin-charge separation.
When $k\rightarrow 0$, the holon band connects to the shadow band, and when $k\rightarrow \pi/2$, it connects to the spinon band.
According to the theoretic analysis~\cite{PhysRevB.45.13156,Sorella_1992}, the spinon bandwidth is $\pi J/2$ when $U\rightarrow\infty$.
With $U/t=8$, the width of the spinon band in Fig.~\ref{Fig.main3}(a) is around $0.75t$, which is close to the analytical estimate $\pi J/2\sim2\pi t^2/U=0.785t$.
Moreover, the velocity of a spinon is slower than that of a holon due to the forward scattering of Coulomb interaction~\cite{QP1D}. 
When the on-site $e$-ph couplings are included, HM becomes HHM.
By integrating out the phonon degrees of freedom, we obtain an effective on-site attraction $-\lambda/(1-\omega^2/\omega_0^2)\sum_{i}\hat{n}_{i,\uparrow}\hat{n}_{i,\downarrow}$~\cite{PhysRevLett.96.156402}. 
The phonon-mediated coupling strength is set as a moderate value $\lambda=2\gamma^2/\omega_0=2t$.
The spectra of the half-filled HHM are shown in Fig.~\ref{Fig.main3}(b). 
Because of phonon effects, the holon band is broadened with reduced spectral weight while the shadow band is almost smeared out. 
In contrast to HM, the holon and shadow bands split in Fig.~\ref{Fig.main4}(c) when $k\rightarrow 0$.
Furthermore, the holon and spinon bands are not connected when $k\rightarrow \pi/2$, where the split value $1.075t$ is close to the phonon frequency $\omega_0=t$.
These features are consistent with Fig.~5 in Ref.~\cite{PhysRevLett.96.156402} using CPT and have a higher resolution.

\subsection{Doped eHHM with $\omega_0/t=1$}

\begin{figure}[tbp]
\centering 
\includegraphics[width=1.0\columnwidth]{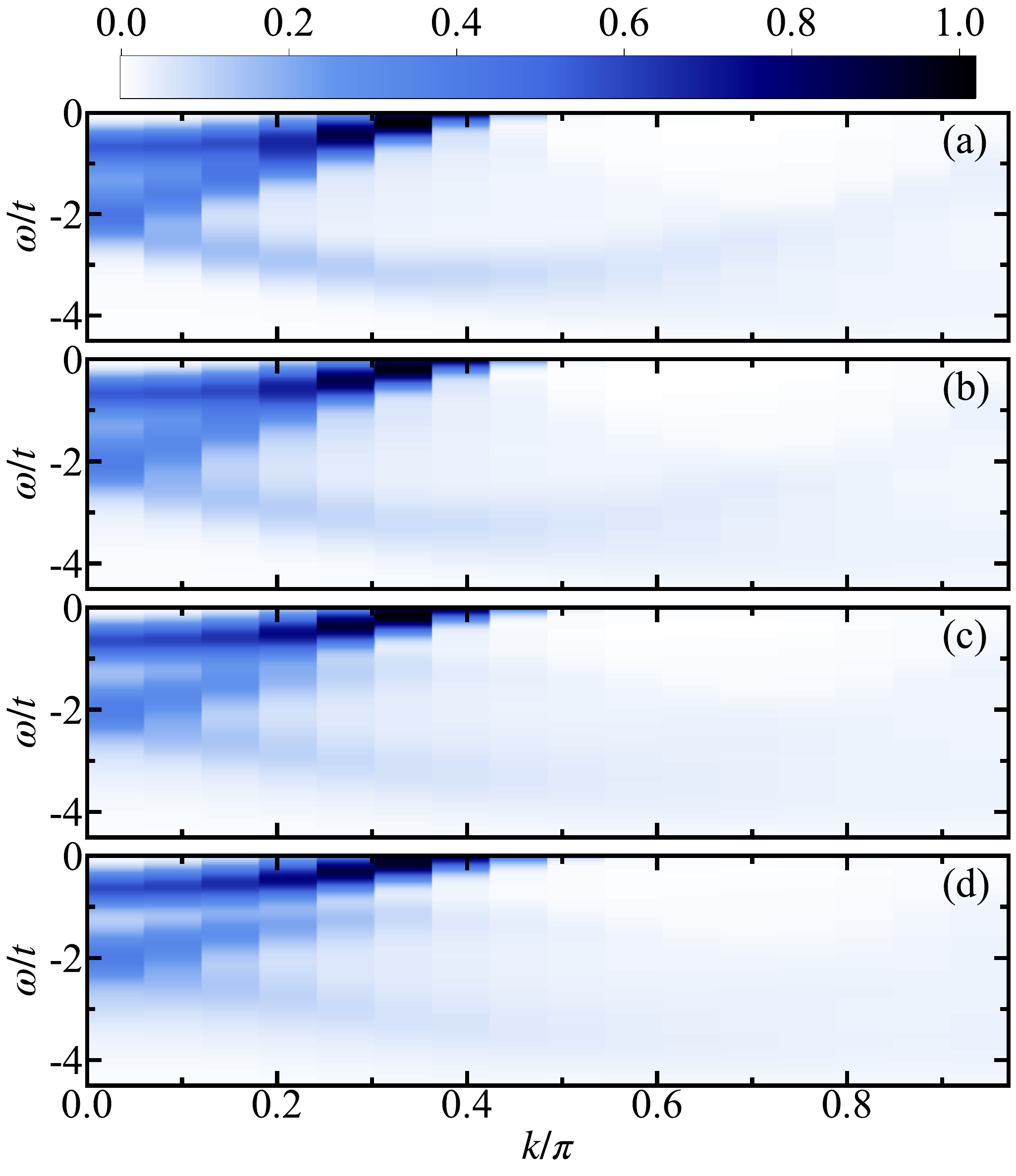}
\caption{
The spectral functions of the doped models with $N=32$, $N_e=24$, $U/t=8$, $N_{\mathrm{C}}=100$, $\omega_0/t=1$, $d_K=30$, $\omega_{1\mathrm{max}}=15$, and $\omega_{2\mathrm{max}}=3$.
(a) HM with $D=100$. 
(b) HHM with $D=200$, $N_p=3$, and $\gamma/t=0.5$. 
(c) eHHM with $D=200$, $N_p=4$, $\gamma/t=0.5$, and $\gamma'=\gamma/\sqrt{5}$.
(d) eHHM with $D=200$, $N_p=4$, $\gamma/t=0.5$, $\gamma'=\gamma/\sqrt{5}$, and $\gamma''=\gamma/\sqrt{17}$.
} \label{Fig.main5}
\end{figure}

\begin{figure}[tbp]
\centering 
\includegraphics[width=1.0\columnwidth]{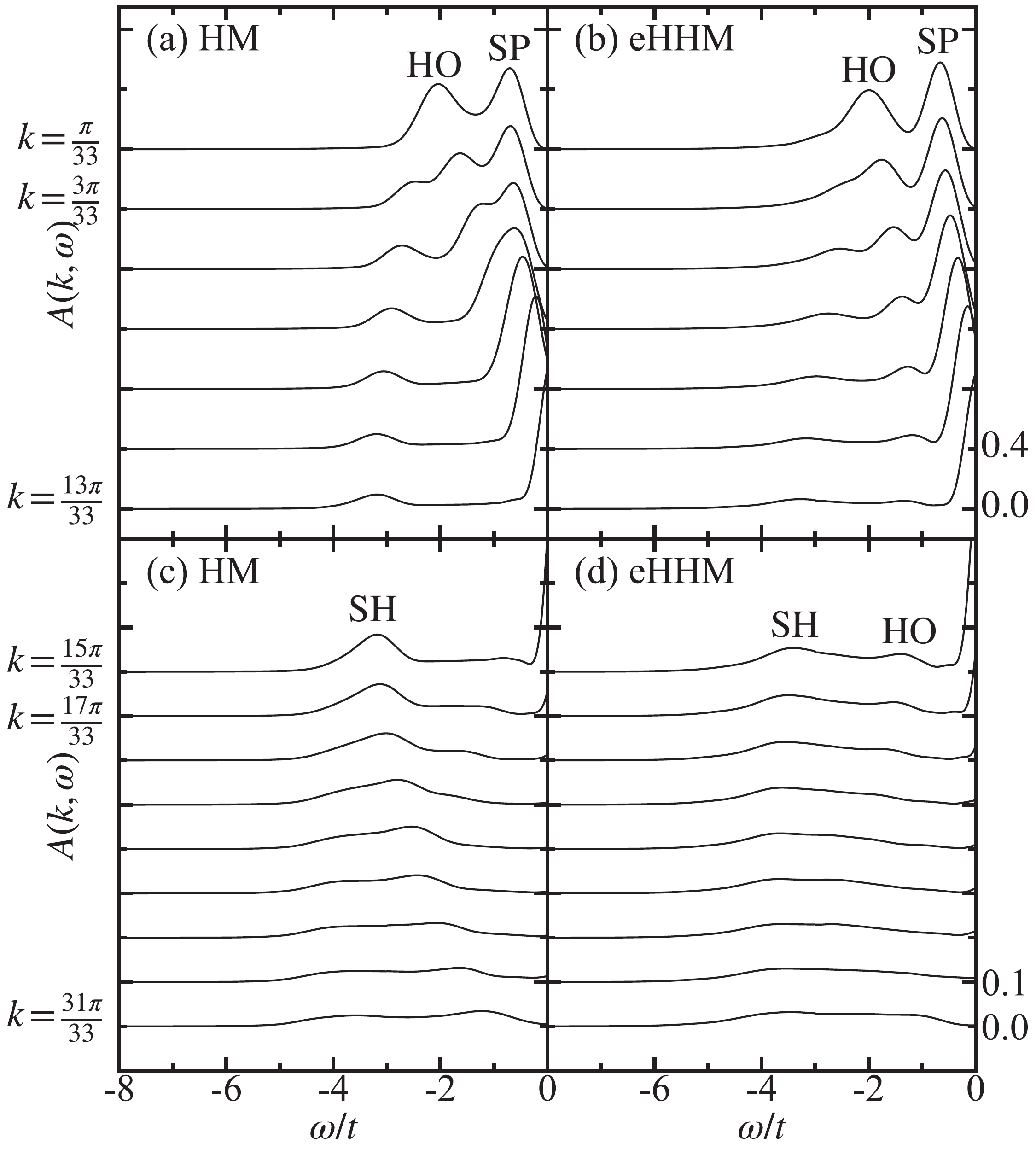} 
\caption{
Energy distribution curves for spectral functions of doped models. 
Here, $k_F=3 \pi/8$.
(a) and (c) are from Fig.~\ref{Fig.main5}(a), while (b) and (d) are from Fig.~\ref{Fig.main5}(d).
} 
\label{Fig.main6} 
\end{figure}

Inspired by experiments on 1D cuprates~\cite{chen2021anomalously,wang2021phonon}, we study the phonon effects on the spectral functions of the doped eHHM with electron density per site $\nu=N_e/N$. 
This model includes the on-site ($\gamma$), NN ($\gamma'$), and NNN ($\gamma''$) $e$-ph couplings, where $\gamma'=\gamma/\sqrt{7}$ and $\gamma''=\gamma/\sqrt{15}$. 
These ratios are based on the approximate geometry distances between apical oxygen and copper atoms in 1D cuprates~\cite{chen2021anomalously,wang2021phonon}.
The chemical potential $\mu$ is chosen to ensure that $E_0(N_e-1)=E_0(N_e+1)$, where $E_0(N_e\pm1)$ represents the ground state energy with $N_e\pm 1$ electrons.
In the thermodynamic limit with $N\rightarrow \infty$, the Fermi level is at $\omega=0$~\cite{PhysRevLett.92.256401}.

We begin by examining the $\nu=3/4$ model with phonon frequency comparable to electron hopping ($\omega_0/t=1$). 
The spectra of the doped HM are shown in Fig.~\ref{Fig.main5}(a), where the charge gap is nearly closed at $k\rightarrow k_F=\nu\pi/2$ due to partial filling and finite broadening. 
Similar to the Mott insulator in Fig.~\ref{Fig.main3}(a), the spin-charge separation is clearly visible.
The holon band merges with the spinon band when $k\rightarrow k_F$ and connects to the shadow band as $k\rightarrow 0$.
Fig.~\ref{Fig.main5}(b) shows the HHM with on-site $e$-ph couplings $\gamma/t=0.5$.
The $e$-ph coupling $\lambda=0.5t$ is relatively weak in our case, and the spectral function in Fig.~\ref{Fig.main5}(b) is almost identical to that in Fig.~\ref{Fig.main5}(a).
To explore the effects of longer-range phonons, we gradually add more extended $e$-ph interactions in Fig.~\ref{Fig.main5}(c) and (d).
In these figures, the spinon band remains nearly unchanged, and the shadow band still connects to the holon band when $k\rightarrow 0$. 
However, when $k\rightarrow k_F$, the spinon and holon bands are separated, and the split value increases with increasing $e$-ph coupling range.
At the same time, the shadow and holon bands are broadened with decreasing spectral weights from Fig.~\ref{Fig.main5}(a) to (d).

The energy distribution curves (EDC) shown in Fig.~\ref{Fig.main6} correspond to Fig.~\ref{Fig.main5} (a) and (d).
In comparison to the doped HM, the extended $e$-ph coupling splits the spinon and holon bands as $k\rightarrow k_F$, causing the holon band to remain present even when $k>k_F$. 
When $k\rightarrow k_F$, the separation between the spinon and holon bands is approximately $1.02t$, which is close to the phonon frequency. 
Additionally, the holon and shadow bands are broadened and their spectral weights are reduced, similar to the behavior of the half-filled HHM.
These observations demonstrate that even though the on-site $e$-ph coupling ($\gamma$) is very weak, the extended $e$-ph couplings ($\gamma',\gamma''$) can significantly affect the spectra.

\subsection{Doped eHHM with $\omega_0/t=0.2$}

\begin{figure}[tbp]
\centering 
\includegraphics[width=1.0\columnwidth]{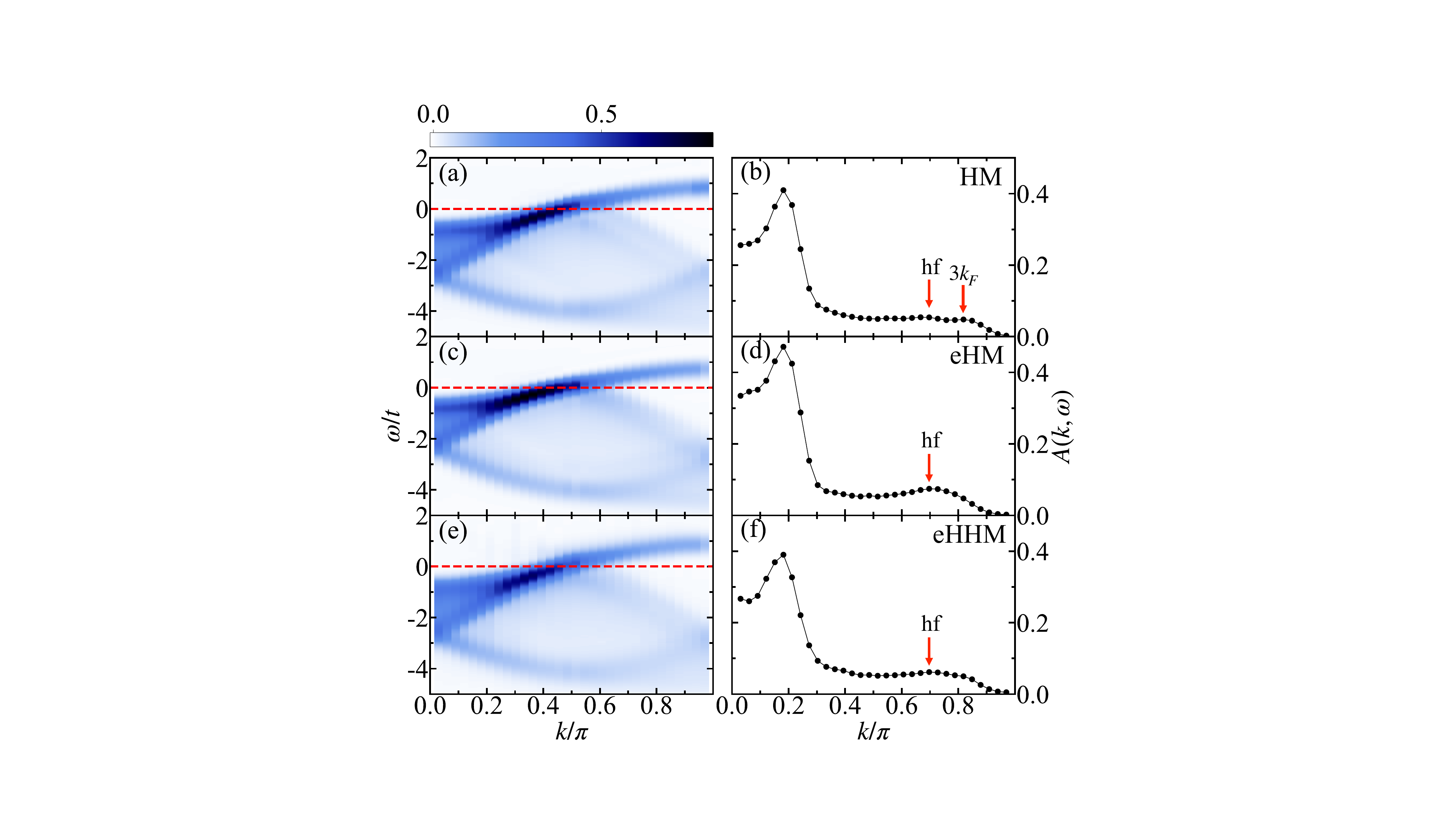}
\caption{
The spectral functions of the doped models with $N=32$, $N_e=30$, $U/t=8$, $N_{\mathrm{C}}=100$, $\omega_0/t=0.2$, $d_K=30$, $\omega_{1\mathrm{max}}=15$, and $\omega_{2\mathrm{max}}=3$. 
The labels hf and $3k_F$ mark the holon-folding band and $3k_F$ band, respectively.
(a) and (b) are for HMs with $D=100$. 
(c) and (d) are for eHMs with $D=100$ and $V=-1$. 
(e) and (f) are for eHHMs with $D=300$, $N_p=4$, $\gamma/t=\sqrt{0.06}$, and $\gamma'=\gamma/\sqrt{5}$.
In (a), (c), and (e), the dashed red line is the Fermi level.
(b), (d), and (f) are the momentum distribution curves (MDC) of (a), (c), and (e), respectively.
} 
\label{Fig.main7}
\end{figure}

Research on 1D cuprates can help us understand the mechanisms behind high-$T_c$ superconductivity.
Recent ARPES experiments on 1D cuprates~\cite{chen2021anomalously} have shown the existence of spin-charge separation. 
Initially, the 1D Hubbard model with $U/t=8$ is used to explain this phenomenon, but the resulting spectral weights are slightly inconsistent with experimental results~\cite{chen2021anomalously}.
Therefore, the extended Hubbard model (eHM) with NN attractive interactions $V\sum_i\hat{n}_i\hat{n}_{i+1}$ has been introduced to improve agreement with experiments~\cite{chen2021anomalously}. 
The origin of this attractive interaction is explored in the 1D eHHM~\cite{wang2021phonon}, which suggests it may be caused by extended $e$-ph couplings.
However, further simulations on the spectra of eHHM are needed to confirm its validity for describing 1D cuprates.

In the following, we simulate the spectra of an 1D eHHM with $N=32$, $N_e=30$, $U/t=8$, $\omega_0/t=0.2$, $\gamma/t=\sqrt{0.06}$, and $\gamma'=\gamma/\sqrt{5}$.
We set the electron density at $\nu=30/32$ (doping level $\delta=1-\nu=6.25\%$) for two reasons. 
First, previous research~\cite{chen2021anomalously} has shown that the lower the doping level $\delta$, the greater the hf band enhancement. 
Second, calculating spectral functions is challenging at high $\nu$ since more phonons are involved in $e$-ph couplings, making it an excellent platform to demonstrate the power of our algorithm.
The phonon frequency is extracted from experiments as $\omega_0/t\sim 0.12$~\cite{chen2021anomalously,wang2021phonon}, so we choose a similar value $\omega_0/t= 0.2$.
To ensure that the effective attractive interaction is strong enough, the $e$-ph coupling parameter $\gamma/t$ must be as large as possible, but it must also avoid being too large, which would result in the system entering phase separation. 
Therefore, we choose a moderate $\gamma/t=\sqrt{0.06}$.

To compare with the results of 1D eHHM, we first calculate the spectra of HM and eHM. 
As shown in Fig.~\ref{Fig.main7}(a), the spin-charge separation is still present in HM, and the hf and $3k_F$ bands are clearly visible.
The hf band folds down directly at $k_F$, whereas the $3k_F$ band goes beyond the Fermi surface and then bends at $2\pi-3k_F$.
To compare the spectral weights of hf band and $3k_F$ band in different models, we plot the momentum distribution curves (MDC) at an energy roughly $t$ above the holon band bottom to ensure that the main peak of the holon band is roughly in the same position \cite{chen2021anomalously}.
According to the MDC in Fig.~\ref{Fig.main7}(b), the spectral weight of the hf band is very close to that of the $3k_F$ band, which is consistent with Fig. 4 in Ref.~\cite{chen2021anomalously}.
Fig.~\ref{Fig.main7}(d) shows an enhanced hf band and a suppressed $3k_F$ band for eHM, which also coincides with Fig. 4 in Ref.~\cite{chen2021anomalously}.
As shown in Fig.~\ref{Fig.main7}(e-f), extended $e$-ph couplings cause the spectral weight of the hf band to be greater than that of the $3k_F$ band, similar to the phenomenon in Fig.~\ref{Fig.main7}(c-d).
This means that extended $e$-ph couplings can indeed induce effective attractive interactions between electrons.

Our study has shown that extended $e$-ph couplings can have a significant impact on spectral functions.
These interactions can be included in the microscopic theory of 1D cuprates.
Additionally, our method has proven to be successful in calculating challenging parameters, indicating its potential as a powerful tool for studying strongly correlated electron-phonon systems.

\subsection{Error and convergence analysis}

\begin{figure*}[tbp]
\centering 
\includegraphics[width=0.62\textwidth]{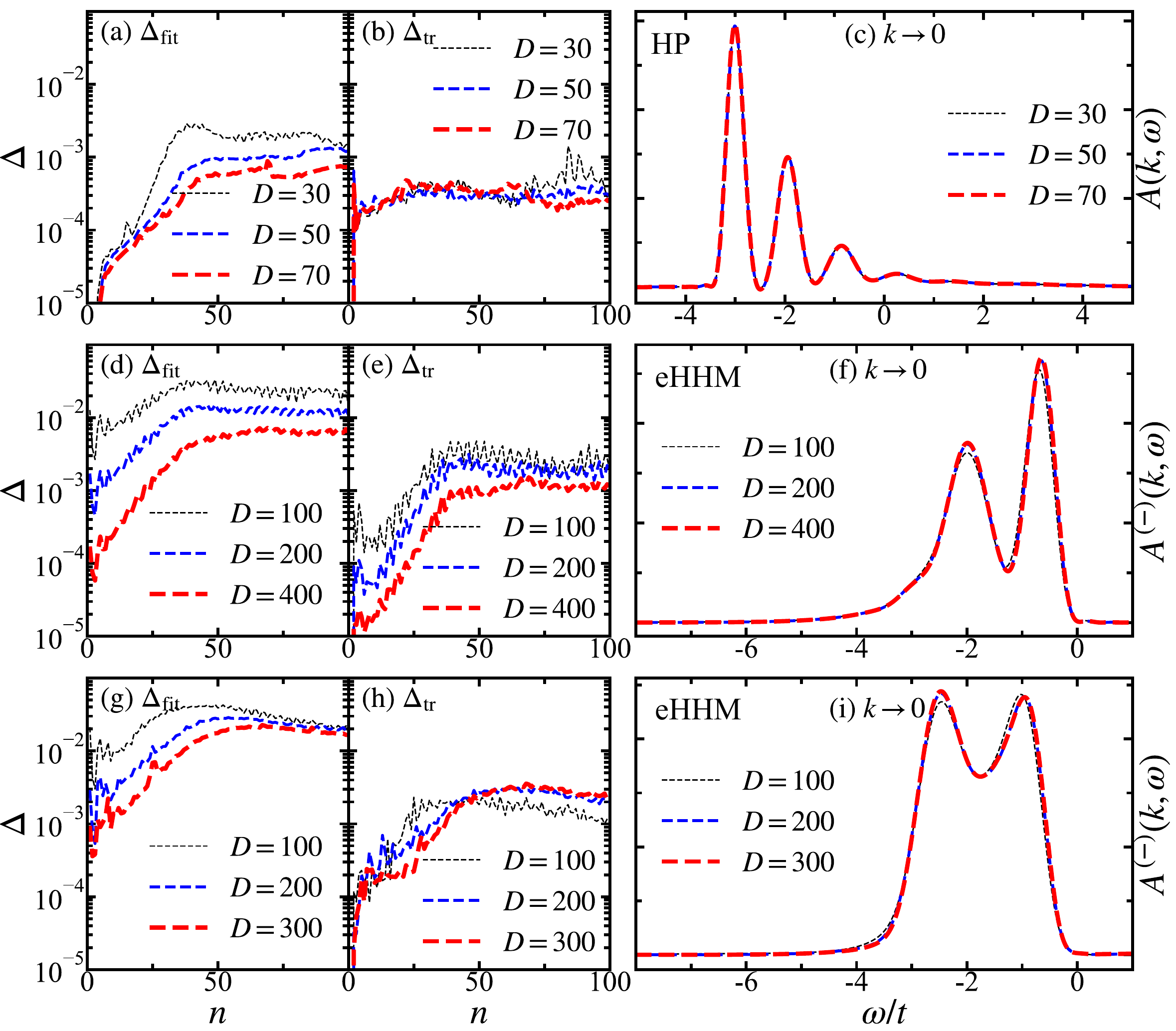}
\caption{
(a-c) The iteration error $\Delta_{\mathrm{fit}}$ (a), truncation-induced state change $\Delta_{\mathrm{tr}}$ (b), and spectra $A(k,\omega)$ at $k\rightarrow 0$ (c) for the Green's function of the Holstein polaron, calculated using the Chebyshev SPS-MPS method with different bond dimensions and parameters as in Fig.~\ref{Fig.main2}.
$n$ is the order of the Chebyshev expansion.
(d-f) $\Delta_{\mathrm{fit}}$, $\Delta_{\mathrm{tr}}$, and $A^{(-)}(k,\omega)$ of eHHM.
Other parameters are identical to those in Fig.~\ref{Fig.main5}(d). 
(g-i) $\Delta_{\mathrm{fit}}$, $\Delta_{\mathrm{tr}}$, and $A^{(-)}(k,\omega)$ of eHHM, with the same parameters as in Fig.~\ref{Fig.main7}(e).
}
\label{Fig.main8} 
\end{figure*}

To check the convergence of our method, we show the fitting errors $\Delta_{\mathrm{fit}}$ and the truncation-induced state changes $\Delta_{\mathrm{tr}}$ with respect to the order of Chebyshev expansion $n$ for various bond dimensions. 
When the phonon frequency is moderate ($\omega_0/t=1$), increasing bond dimension significantly reduces $\Delta_{\mathrm{fit}}$, as seen in Fig.~\ref{Fig.main8}(a) and (d). 
$\Delta_{\mathrm{fit}}$ for the Green's function of the Holstein polaron with $D=70$ is less than $10^{-3}$, and for the photoemission spectra of the doped eHHM with $D=400$ is less than $10^{-2}$. 
However, increasing bond dimension only gradually decreases $\Delta_{\mathrm{tr}}$ in Fig.~\ref{Fig.main8}(b) and (e), indicating that energy truncation is still a limiting factor~\cite{PhysRevB.83.195115}. 
In the adiabatic limit ($\omega_0/t=0.2$) shown in Fig.~{\ref{Fig.main8}}(g), we find that increasing bond dimension only gradually reduces $\Delta_{\mathrm{fit}}$, especially when $n>75$, and $\Delta_{\mathrm{tr}}$ does not decrease.
Nevertheless, the spectral functions are almost the same with increasing bond dimensions in Fig.~\ref{Fig.main8}(c), (f), and (i), indicating that the Chebyshev pseudosite MPSs with small bond dimensions already contain key features of the excitation spectra.

\section{Conclusion}
\label{section4}

In summary, we propose a simple and efficient approach for calculating the spectral functions of $e$-ph coupling systems by combing the $U(1)$-symmetric pseudosite MPS and the Chebyshev MPS approach. 
 We map a real boson with $2^{N_p}$ states to $N_p$ hard-core boson pseudosites by reshaping and decomposing the bosonic MPO.
We demonstrate the effectiveness of this method by numerically solving the Green's function of the Holstein polaron and the spectra of the half-filled HHM in the Mott insulator phase. 
We show that key aspects of the excitation spectra can be captured at a modest computational cost. 
Furthermore, we investigate the doped eHHM with on-site, NN, and NNN $e$-ph interactions and find that extended $e$-ph couplings can significantly influence the spectra. 
These interactions may play an important role in the physics of high-temperature superconductors.  

Our approach is compatible with Fermi-Bose Hamiltonians where pseudosite DMRG performs well.
It works for Hamiltonians with long-range $e$-ph couplings and may be extended to higher dimensions by combining pseudosite methods with projected entangled-pair states (PEPS)~\cite{Verstraete2004,RevModPhys.93.045003}.
Given its flexibility and efficiency, we believe our method will be a valuable and essential tool for investigating strongly correlated electron-phonon coupling systems.

\section{Note added}
After finishing the first version of this article, we notice that Tang \textit{et al.} use tDMRG method to calculate the spectral functions of the 1D eHHM for $\omega_0/t=0.2$~\cite{tang2022traces}. 
Our results qualitatively agree with theirs when the doping level is low.
However, due to the maximum real-time propagation of the finite size system used in their study, the energy resolution of their spectrum function is relatively low~\cite{tang2022traces}. 
Furthermore, we also notice a recent study of the 1D eHM spectrum functions related to 1D cuprates~\cite{wang2022spectral}, but this study does not include phonon degrees of freedom. 

\section{Acknowledgement}

Pei-Yuan Zhao thanks Zheng-Tao Xu for useful discussions. 
This work is supported by the National Natural Science Foundation of China (NSFC) (Grant No. 12174214 and No. 92065205), the National Key R\&D Program of China (Grant No. 2018YFA0306504), and the Innovation Program for Quantum Science and Technology (Grant No. 2021ZD0302100).

\appendix
\section{The explicit pseudosite MPOs for Hubbard Holstein model}
\label{appendixA}

In this section, we present the explicit form of the $U(1)$-symmetric pseudosite MPO for the Hamiltonian of HHM.

The $i$-th lattice site basis is represented as $|\psi_i\rangle_F\otimes|n_{i}\rangle_B$, where $|\psi_i\rangle_F$ and $|n_{i}\rangle_B$ are the fermion and boson bases, respectively. 
$|\psi_i\rangle_F$ can represent the vacuum state $|\mathrm{vac}\rangle$, the double occupancy state $\hat{c}_{i,\uparrow}^{\dagger}\hat{c}_{i,\downarrow}^{\dagger}|\mathrm{vac}\rangle$, the single occupancy state $\hat{c}_{i,\uparrow}^{\dagger}|\mathrm{vac}\rangle$, or the single occupancy state $\hat{c}_{i,\downarrow}^{\dagger}|\mathrm{vac}\rangle$. 
$|n_{i}\rangle_B$ is an eigenvector of the particle number operator for bosons, i.e. $\hat{a}_i^{\dagger}\hat{a}_i|n_{i}\rangle_B=n_{b,i}|n_{i}\rangle_B$. 

The Jordan-Wigner transformation for the fermionic creation and annihilation operators is 
\begin{align}
\hat{c}^{\dagger}_{i\uparrow}&=(-1)^{\sum_{j<i}(\hat{n}_{j,\uparrow}+\hat{n}_{j\downarrow})}\hat{S}^{+}_{i,\uparrow},\nonumber\\
\hat{c}_{i\uparrow}&=(-1)^{\sum_{j<i}(\hat{n}_{j,\uparrow}+\hat{n}_{j\downarrow})}\hat{S}^{-}_{i,\uparrow},\nonumber\\
\hat{c}^{\dagger}_{i\downarrow}&=(-1)^{\sum_{j<i}(\hat{n}_{j,\uparrow}+\hat{n}_{j\downarrow})}(-1)^{\hat{n}_{i,\uparrow}}\hat{S}^{+}_{i,\downarrow},\nonumber\\
\hat{c}_{i\downarrow}&=(-1)^{\sum_{j<i}(\hat{n}_{j,\uparrow}+\hat{n}_{j\downarrow})}(-1)^{\hat{n}_{i,\uparrow}}\hat{S}^{-}_{i,\downarrow},\nonumber\\
\label{EqA5}
\end{align}
where $\hat{n}_{i,\sigma}=\hat{c}^{\dagger}_{i,\sigma}\hat{c}_{i,\sigma}=\hat{S}^+_{i,\sigma}\hat{S}^-_{i,\sigma}$ and $\sigma=\uparrow,\downarrow$.
Using Eq.~(\ref{EqA5}), the hopping terms are represented as
\begin{align}
\hat{c}_{i,\uparrow}^{\dagger}\hat{c}_{i+1,\uparrow}&=\hat{S}_{i,\uparrow}^{+}(-1)^{\hat{n}_{i,\uparrow}+\hat{n}_{i,\downarrow}}\hat{S}_{i+1,\uparrow}^{-},\nonumber\\
c_{i,\downarrow}^{\dagger}c_{i+1,\downarrow}&=\hat{S}_{i,\downarrow}^{+}(-1)^{\hat{n}_{i,\downarrow}+\hat{n}_{i+1,\uparrow}}\hat{S}_{i+1,\downarrow}^{-}.
\end{align}

Next, we provide the $U(1)$-symmetric MPO.
For simplicity, we first define
\begin{align}
\hat{M}_1(i)&=\hat{S}_{i,\uparrow}^{+}(-1)^{\hat{n}_{i,\uparrow}+\hat{n}_{i,\downarrow}},\ \hat{M}_2(i)=\hat{S}_{i,\uparrow}^{-},\nonumber\\
\hat{M}_3(i)&=\hat{S}_{i,\downarrow}^+(-1)^{\hat{n}_{i,\downarrow}},\ \hat{M}_4(i)=(-1)^{\hat{n}_{i,\uparrow}}\hat{S}_{i,\downarrow}^{-},
\end{align}
so that
\begin{align}
\hat{c}_{i,\uparrow}^{\dagger}\hat{c}_{i+1,\uparrow}&=\hat{M}_1(i)\hat{M}_2(i+1),\nonumber\\
\hat{c}_{i,\downarrow}^{\dagger}\hat{c}_{i+1,\downarrow}&=\hat{M}_3(i)\hat{M}_4(i+1).
\end{align}
Since the MPO of the HHM Hamiltonian is translational invariant, we ignore the index $i$ in Eq.~(\ref{EqA9})-Eq.~(\ref{EqA16}). 
The MPO of the fermionic degrees of freedom is represented as
\begin{align}
&W_F=\nonumber\\
&\left(
\begin{array}{ccc|cc|cc}
\hat{\mathbb{I}}_F&0&0&0&0&0&0\\
0&0&0&0&0&0&0\\
U\hat{n}_{\uparrow}\hat{n}_{\downarrow}-\mu \hat{n}&\gamma \hat{n}&\hat{\mathbb{I}}_F&-t\hat{M}_1&-t\hat{M}_3&-t\hat{M}_1^{\dagger}&-t\hat{M}_3^{\dagger}\\
\hline
M_2&0&0&0&0&0&0\\
M_4&0&0&0&0&0&0\\
\hline
M_2^{\dagger}&0&0&0&0&0&0\\
M_4^{\dagger}&0&0&0&0&0&0\\
\end{array}\right),
\label{EqA9}
\end{align}
where $\hat{\mathbb{I}}_F$ is a $4 \times 4$ identity operator.
The MPO of the bosonic degrees of freedom is
\begin{align}
W_B=\left(\begin{array}{ccc|cc|cc}
\hat{\mathbb{I}}_B&0&0&0&0&0&0\\
\hat{a}^{\dagger}+\hat{a}&0&0&0&0&0&0\\
\omega_0\hat{a}^{\dagger}\hat{a}&0&\hat{\mathbb{I}}_B&0&0&0&0\\
\hline
0&0&0&\hat{\mathbb{I}}_B&0&0&0\\
0&0&0&0&\hat{\mathbb{I}}_B&0&0\\
\hline
0&0&0&0&0&\hat{\mathbb{I}}_B&0\\
0&0&0&0&0&0&\hat{\mathbb{I}}_B\\
\end{array}\right),
\end{align}
where $\hat{\mathbb{I}}_B$ is a $2^{N_p}\times2^{N_p}$ identity operator and $N_p$ is the number of pseudosites. 
$W_B$ has a block-sparse form
\begin{align}
W_B=\left(\begin{array}{c|c|c}
(W_B)_{0,0}&&\\
\hline
&(W_B)_{-1,-1}&\\
\hline
&&(W_B)_{1,1}\\
\end{array}\right),
\end{align}
where the labels of blocks are quantum numbers for the virtual indices of $W_B$.

After that, the bosonic MPO $W_B$ is decomposed into a set of pseudosite MPOs 
\begin{align}
W_B=W_{p1}W_{p2}\cdots W_{pN_p}.
\end{align}
This factorization is performed block by block. 
$(W_B)_{0,0}$ is decomposed by SVD~\cite{SCHOLLWOCK201196,PhysRevA.81.062337,PhysRevB.95.035129}
\begin{align}
(W_B)_{0,0}=(W_{p1})_{0,0}(W_{p2})_{0,0}\cdots (W_{pN_p})_{0,0}
\end{align}
and $(W_B)_{\pm1,\pm1}$ is substituted by a set of identity MPOs
\begin{align}
(W_B)_{\pm1,\pm1}&=W_{I_p1}W_{I_p2}\cdots W_{I_pN_p},\nonumber \\
W_{I_pj}&=\begin{pmatrix}\hat{\mathbb{I}}_p&0\\0&\hat{\mathbb{I}}_p\end{pmatrix},\ \ 1\leq j\leq N_p,
\end{align}
where $\hat{\mathbb{I}}_p$ is a $2 \times 2$ identity operator for the bosonic pseudosite.
The pseudosite MPO is further represented as
\begin{align}
W_{pj}=\left(\begin{array}{c|c|c}
(W_{p_j})_{0,0}&&\\
\hline
&W_{I_pj}&\\
\hline
&&W_{I_pj}\\
\end{array}\right).
\end{align}
Finally, $W_F$ and $W_{pj}$ constitute the MPO for one lattice site
\begin{align}
\mathbb{W}=W_FW_{p1}W_{p2}\cdots W_{pN_p}.
\label{EqA16}
\end{align}
 
\bibliography{apssamp}
\end{document}